\documentclass[natbib]{iau}

\usepackage{amsmath}
\usepackage{graphicx}
\usepackage{multirow}
\usepackage{mhchem}

\newcommand{\msunyr}{M$_\odot$\, yr$^{-1}$}
\newcommand{\kms}{km\, s$^{-1}$}

\begin{document}

\lefttitle{Marie Van de Sande}
\righttitle{Chemical complexity and dust formation around evolved stars}

\jnlPage{1}{7}
\jnlDoiYr{2021}
\doival{10.1017/xxxxx}

\aopheadtitle{Proceedings IAU Symposium}
\editors{E. Bergin, P. Caselli, J. J{\o}rgensen, eds.}

\title{Chemical complexity and dust formation around evolved stars}

\author{Marie Van de Sande}
\affiliation{Leiden Observatory, Leiden University, P.O. Box 9513, 2300 RA Leiden, The Netherlands}
\affiliation{School of Physics and Astronomy, University of Leeds, Leeds LS2 9JT, UK}

\begin{abstract}
The outflows of asymptotic giant branch (AGB) stars are rich astrochemical laboratories, hosting different chemical regimes: from non-equilibrium chemistry close to the star, to dust formation further out, and finally photochemistry in the outer regions.
Chemistry is crucial for understanding the driving mechanism and dynamics of the outflow, as it is the small-scale chemical process of dust formation that launches the large-scale stellar outflow.
However, exactly how dust condenses from the gas phase and grows is still unknown: an astrochemical problem with consequences for stellar evolution.
Disagreements between observations and the predictions of chemical models drive the development of these models, helping to understand the link between dynamics and chemistry and paving the way to a 3D hydrochemical model.
\end{abstract}


\maketitle

\section{Introduction}

During the asymptotic giant branch (AGB) phase, stars with an initial mass between 0.8 and 8 M$_\odot$ are stripped of their outer layers by a stellar outflow or wind.
This mass loss is efficient, with mass-loss rates between $10^{-8}$ and $10^{-4}$ \msunyr, and creates an extended circumstellar envelope (CSE).
It is this process of mass loss that determines the star’s remaining lifetime, rather than exhaustion of nuclear fuel in its core.
The stellar outflow is thought to be driven by a two-step mechanism: stellar pulsations facilitate dust formation, which then launches a dust-driven wind \citep{Habing2003,Hofner2018}. 
Thanks to their outflows, AGB stars are important contributors to the chemical enrichment of the interstellar medium (ISM): about 80\% of gas in the ISM originates from AGB stars \citep{Tielens2005} and they produce about 70\% of the total stardust production rate, making them the main contributors of stellar dust to the ISM \citep{Zhukovska2013}.
This newly formed stellar dust forms the starting point of the chemical and physical evolution of dust in the ISM.

The CSEs of AGB stars are rich astrochemical laboratories: more than 100 molecules have been detected in AGB outflows so far (roughly half of all interstellar molecules), along with some 15 different types of dust grains \citep{Decin2021}.
Understanding their chemistry is crucial to stellar evolution, as the large-scale dynamical outflow is driven by the small-scale chemical process of dust formation.
AGB chemistry is not just important to stellar death, but also to stellar birth since they enrich the ISM with gas and newly formed dust -- the building blocks of the next generation of stars and planets.

In Sect. \ref{sect:CSE}, I expand on the types of chemistry chemistry throughout the outflow and the different chemical models used to understand it.
I then discuss two key topics in the chemistry of AGB outflows: understanding dust formation and its properties in Sect. \ref{sect:dust} and unravelling disagreements between gas-phase observations and chemical model predictions in Sect. \ref{sect:compl}.
Finally, Sect. \ref{sect:discussion} highlights some modelling and observational necessities to resolve these issues.

\section{The circumstellar envelope}		\label{sect:CSE}

AGB outflows are unique environments that host different types of chemistry, including dust formation.
Moreover, as its an outflow, these different regimes are more easily probed by observations than in other astrochemical laboratories, such as protoplanetary disks or dark clouds.
Sect. \ref{subsect:CSE:struct} expands on the different types of chemistry throughout the outflow, Sect. \ref{subsect:CSE:models} describes the chemical models used to describe these different chemical regimes.

\begin{figure}[t]
	\includegraphics[width=1\textwidth]{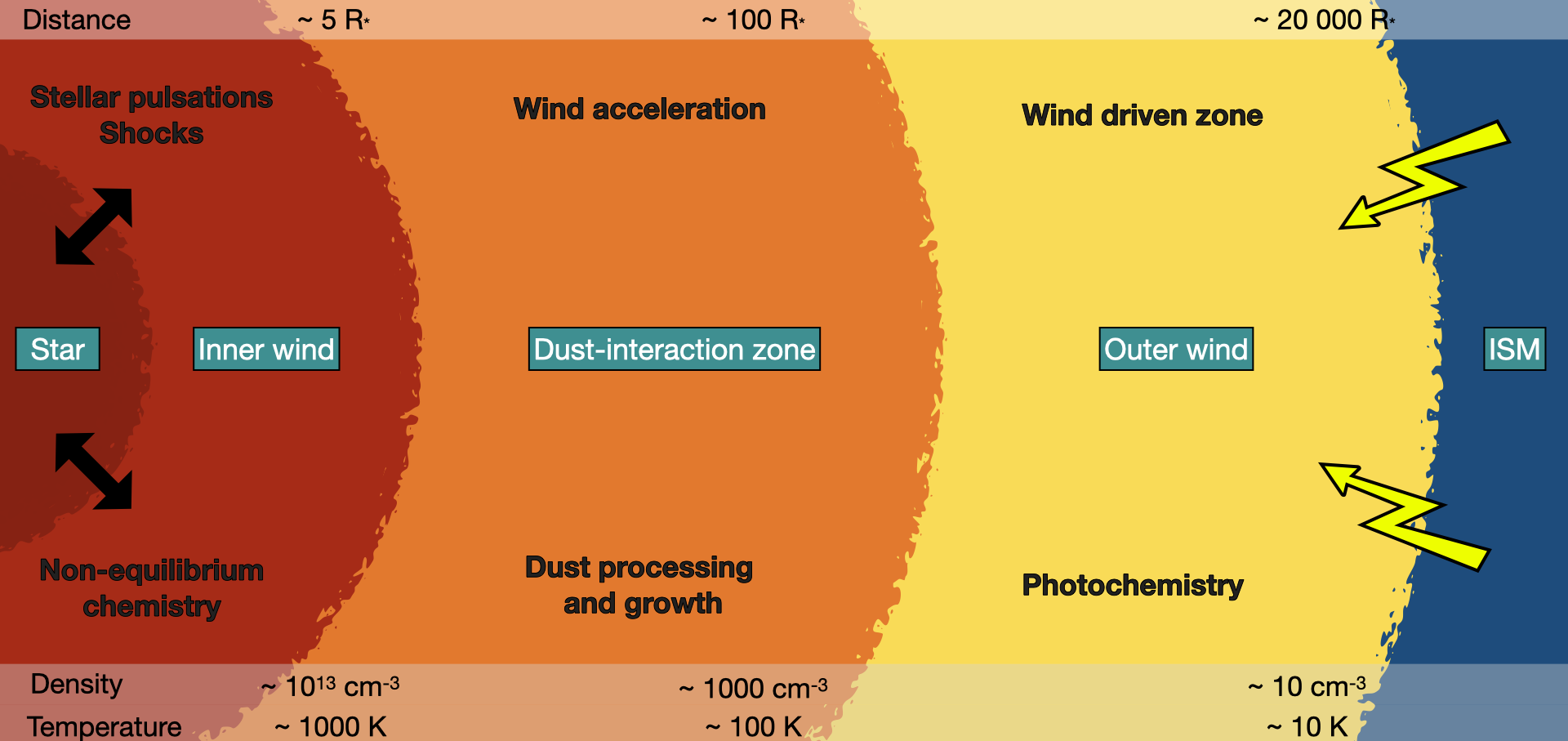}
    \caption{Schematic overview of a single AGB star  and its spherically symmetric CSE.
    The arrows on the stellar surface (left) represent stellar pulsations. 
    The lightning bolts (right) represent interstellar UV photons penetrating the lower-density outer wind.
    Dynamical processes are indicated in the top half of the figure, chemical processes in the bottom half.
    Typical sizes are given at the top, the gradients in density and temperature throughout the outflow are given at the bottom.
    }
    \label{fig:outflow}
\end{figure}

\subsection{Structure and composition}			\label{subsect:CSE:struct}

Fig. \ref{fig:outflow} shows a schematic overview of an AGB star and its outflow. 
The AGB star itself is large, with a stellar radius of $\approx 1$ au. 
Its CSE extends from the stellar surface to roughly 20,000 R$_*$, where it merges with the ISM.
Thanks to the large gradients in density and temperature, different types of chemistry are present throughout the outflow.
The outflow can be divided into three main regions, each characterised by closely-linked chemical and dynamical processes
\citep{Habing2003,Hofner2018,Decin2021}:
\begin{enumerate}
	\item The \emph{inner wind} extends from the stellar surface to some 5 R$_*$.
	At the high temperatures and densities of the stellar photosphere, thermodynamic equilibrium (TE) is a good assumption. 
	However, shocks caused by stellar pulsations take the chemistry out of TE, breaking up most of the stable CO molecule.
	\item In the \emph{intermediate wind} or \emph{dust-interaction zone}, which ranges between $\sim 5 - 100$ R$_*$, the physical conditions allow for solid dust grains to condense from the gas phase.	
	Radiation pressure onto the dust grains accelerates material away from the star, launching the large-scale outflow.
	\item In the \emph{outer wind}, the outflow has reached its terminal expansion velocity which ranges from 3 to 30 \kms, with the majority between 5 and 15 \kms. 
	Because of the lower densities, the chemistry in this region is dominated by photochemistry initiated by interstellar UV photons.
\end{enumerate}

The chemical composition of the outflow depends on the star's elemental carbon-to-oxygen ratio, which is linked to its age and initial mass.
Carbon-rich (C-rich) stars have C/O $> 1$ (initial mass between 1.5 and 4 M$_\odot$) produce amorphous carbon and SiC grains and host a C-rich gas-phase chemistry, with molecules such as \ce{C2H2}, HCN, and carbon-chains abundantly present. 
Oxygen-rich (O-rich) stars have C/O $<1$ and produce silicates and metal oxides and host an O-rich gas-phase chemistry, with molecules such as \ce{H2O}, SiO, and SO abundantly present \citep{Habing2003,Decin2021}.
C-rich stars also produce polycyclic aromatic hydrocarbons (PAHs), which are thought to be the origin of the elusive diffuse interstellar bands \citep{Tielens2013}.
Table \ref{table:parents} lists (average) observed abundances of molecules in the inner winds of C-rich and O-rich outflows, as compiled from observations by \citet{Agundez2020}.
S-type stars are transitioning from C-rich to O-rich chemistry and have C/O $\approx 1$.

\begin{table}[t]
	\caption{Parent species and their abundances relative to \ce{H2} for C-rich and O-rich outflows.
	Abundances are derived from observations, as compiled by \citet{Agundez2020}. 
	} 
    \centering
    {\tablefont \begin{tabular}{l r c  l r }
    \hline  
    \multicolumn{2}{c}{Carbon-rich} && \multicolumn{2}{c}{Oxygen-rich}  \\  
    \cline{1-2} \cline{4-5} 
    \noalign{\smallskip}
    Species & Abundance & & Species & Abundance \\
    \cline{1-2} \cline{4-5} 
    \noalign{\smallskip}
    He		&  0.17				& & He		& 0.17  \\
    CO		& $8.00\times10^{-4}$	& & CO		& $3.00 \times 10^{-4}$  \\
    N$_2$		& $4.00 \times 10^{-5}$	& & H$_2$O	& $2.15 \times 10^{-4}$  \\
    CH$_4$	& $3.50 \times 10^{-6}$	& & N$_2$ 	& $4.00 \times 10^{-5}$  \\ 
    H$_2$O	& $2.55 \times 10^{-6}$	& & SiO 	& $2.71 \times 10^{-5}$  \\ 
    SiC$_2$	& $1.87 \times 10^{-5}$	& & H$_2$S 	& $1.75 \times 10^{-5}$  \\
    CS		& $1.06 \times 10^{-5}$	& & SO$_2$ 	& $3.72 \times 10^{-6}$  \\
    C$_2$H$_2$& $4.38 \times 10^{-5}$	& & SO 		& $3.06 \times 10^{-6}$  \\
    HCN		& $4.09 \times 10^{-5}$	& & SiS 		& $9.53 \times 10^{-7}$  \\
    SiS   		& $5.98 \times 10^{-6}$	& & NH$_3$ 	& $6.25 \times 10^{-7}$  \\ 
    SiO 		& $5.02 \times 10^{-6}$	& & CO$_2$ 	& $3.00 \times 10^{-7}$  \\   
    HCl		& $3.25 \times 10^{-7}$	& & HCN 	& $2.59 \times 10^{-7}$  \\  
    C$_2$H$_4$& $6.85 \times 10^{-8}$	& & PO 		& $7.75 \times 10^{-8}$  \\ 
    NH$_3$	& $6.00 \times 10^{-8}$	& & CS 		& $5.57 \times 10^{-8}$  \\
    HCP		& $2.50 \times 10^{-8}$	& & PN 		& $1.50 \times 10^{-8}$  \\
    HF    		& $1.70 \times 10^{-8}$	& &  HCl		& $1.00 \times 10^{-8}$  \\  
    H$_2$S	& $4.00 \times 10^{-9}$	& & 	HF	& $1.00 \times 10^{-8}$  \\    
    \hline 
    \end{tabular} }%
    \label{table:parents}    
\end{table}


Observations continue to reveal asymmetrical structures within the outflow.
Large-scale structures, such as spirals \cite[e.g.,][]{Mauron2006,Maercker2012} and disks \cite[e.g.,][]{Kervella2016}, and small-scale density inhomogeneities \cite[e.g.,][]{Khouri2016,Agundez2017,VelillaPrieto2023} are ubiquitously observed within CSEs.
Binary interactions with (sub)stellar companions are thought to be the driving mechanism behind these large-scale asymmetries \cite[e.g.,][]{Ramstedt2017,Decin2020}.
The intricate and complex shapes of planetary nebulae \cite[e.g.,][]{Balick2002}, a later evolutionary phase, hence might have their origins in the AGB phase.
Small-scale asymmetries in the inner wind  are likely caused by the convective motions of the star's outer layers \cite[e.g.,][]{Freytag2017,Hofner2019} and can impact the structure of the outflow beyond \cite[e.g.,][]{VelillaPrieto2023}.

\subsection{Chemical models}			\label{subsect:CSE:models}

Chemical models of AGB outflows can be split into three main groups, geared towards the chemistry in each region.
\emph{Inner wind models} deal with the non-TE aspect of the chemistry in this region and typically do not include photochemistry.
The effect of pulsations is included in a semi-analytical \cite[e.g.,][]{Cherchneff2006,Cherchneff2012} or hydrodynamical way \citep{Boulangier2019a}.
These models follow the gas-phase chemistry up to the point of dust formation, predicting possible seed particles \cite[e.g.,][]{Boulangier2019b} and the gas-phase signatures of non-TE chemistry \cite[e.g.,][]{Gobrecht2016}.
\emph{Outer wind models} focus on the rich gas-phase photochemistry in this region, using power laws to describe the density and temperature profiles \cite[e.g.,][]{Huggins1982,Millar1994,Li2016}. 
Both inner and outer wind models typically include gas-phase chemistry only.
The chemical validity of an outer wind model can be extended into the \emph{dust-interaction zone} by including dust-gas interactions and grain-surface chemistry \citep{VandeSande2019b,VandeSande2020,VandeSande2021}. 
The species listed in Table \ref{table:parents} can be used as parent species, i.e., species that are present at the start of a dust-interaction  or outer wind chemical kinetics model.

Chemical models generally assume spherical symmetry.
The effect of \emph{small-scale inhomogeneities or clumps} on the radiation field throughout the outflow can be included by allowing a certain fraction of photons to reach the inner wind \citep{Agundez2010} or by using the mathematical porosity formalism, which accounts for the effect of such clumps on both the radiation field and the local overdensities within the clumps \citep{VandeSande2018}.
The effects of \emph{internal UV photons}, either from the AGB star itself \citep{VandeSande2019a} or from a nearby stellar companion \citep{VandeSande2022}, can also be included. 
In these models, photochemistry already starts in the dense inner wind, rather than be restricted to the tenuous outer wind where it is initiated by interstellar UV photons.
This can have profound effects on the chemistry throughout the outflow, depending on the balance between photoreactions and two-body reactions.
Such models can be combined with a porous outflow as well as dust-gas chemistry \citep{VandeSande2023}.
Finally, several chemical models are dedicated to the density-enhanced shells in the famous C-rich outflow of IRC+10216 \citep{Brown2003,Cordiner2009,Agundez2017}.

\section{Dust formation and its properties}			\label{sect:dust}

Exactly how dust is formed and grows in AGB outflows is still unknown, despite its importance to driving the outflow. 
As a result, the precise composition and grain size distribution of the stellar dust enriching the ISM are also not fully known.
Sect. \ref{subsect:dust:formation} describes different approaches to study dust formation.
Sect. \ref{subsect:dust:properties} discusses how chemical kinetics models can help constrain the dust's properties.

\subsection{Dust formation}			\label{subsect:dust:formation}

Dust formation in C-rich outflows can be studied using a laboratory approach.
The \emph{Stardust} machine of the \textsc{Nanocosmos} team (PIs Cernicharo, Joblin, Martín-Gago)\footnote{\url{https://nanocosmos.iff.csic.es/}} is specifically built to reproduce the physical conditions in AGB outflows, simulating dust nucleation and its interaction with the gas phase \citep{Martinez2018,Martinez2020}.
When creating a more realistic C-rich AGB environment by adding \ce{C2H2}, there was efficient formation of carbon chains, benzene and PAHs, and both the amount of dust and its size distribution changed \citep{Santoro2020}.
Such an in-situ approach allows for a controlled investigation of the dust nucleation and growth processes along with the products enriching the ISM.

In the absence of a laboratory set-up, a combination of observations and theoretical modelling needs to be used for O-rich outflows.
A major issue for O-rich dust is that its seed particles, i.e., the molecules that will form the first gas-phase clusters, are unknown.
Aluminium oxide clusters ((\ce{Al2O3})$_n$) appear to be good candidates because of the retrieved abundances of AlO and AlOH, its direct precursors, close to the high mass-loss rate IK Tau and low mass-loss rate R Dor \citep{Decin2017,Danilovich2020}.
However, models using non-equilibrium dust nucleation can only efficiently form (\ce{Al2O3})$_n$ clusters close to the star if the monomer is already present \citep{Boulangier2019b}.
Since dust is observed to be present close to the star \citep{Norris2012,Khouri2016}, there is a large discrepancy between theory and observations.
This indicates that perhaps cluster formation pathways are missing in the model, or that the current reaction rates involving Al- and O-bearing molecules are not accurate, or both. 
To relieve these tensions and make a more complete nucleation network, quantum chemical calculations of stable clusters and their formation pathways are necessary. 
This has been done for several species, such as silicates \citep{Goumans2012}, calcium titanite \citep{Plane2013}, aluminium oxides \citep{Gobrecht2022}, and titanium oxides \citep{Sindel2023}.
Such a  bottom-up kinetic approach contrasts with the commonly used top-down classical nucleation method of dust formation \cite[e.g.,][]{DellAgli2017}, where seed particles are assumed to have the same properties as the bulk dust grain.

Another example highlighting the interplay between observations and theory are AlCl and AlF around the S-type star W Aql. 
Similar to AlO and AlOH around IK Tau and R Dor, \citet{Danilovich2021} found that these halides are only abundantly present close to the star.
Reactions of Al, AlO, and AlOH with HCl and HF to form AlCl and AlF were estimated using rate theory and added to a gas-phase reaction network. 
While the chemical model was not able to reproduce the observed abundances, it does indicate that AlF is involved in the dust-formation process.
The small extent of AlCl points towards another, as of yet unexplained, chemical or physical factor. 
This study shows how observations guide chemical model development, and how these chemical models in turn can help understand the underlying dynamics or point out missing physics and chemistry.

\subsection{Properties and processing of dust}			\label{subsect:dust:properties}

While dust formation is not understood, dust is abundantly produced and present in the outflow \cite[e.g.,][]{Heras2005}.
Molecular line observations have long pointed towards the effects of dust-gas interactions. 
In high-density O-rich outflows, the abundance of SiO and SiS decreases before its photodissociation, suggesting depletion onto dust grains \citep{Bujarrabal1989,Decin2010a,Verbena2019}.
Looking at larger samples of stars, the abundance of the refractory molecules SiC, SiO, and SiS (tentatively) decreases with increasing outflow density, indicating their involvement in dust formation \citep{GonzalezDelgado2003,Massalkhi2019,Massalkhi2020}. 
Finally, \ce{H2O} ice was detected around OH/IR stars (O-rich stars with high mass-loss rates and high extinction at optical wavelengths, \citealt{Sylvester1999}).

Chemical kinetics models that include dust-gas interactions describe the dust's influence on the abundances of gas-phase species.
In high-density outflows, parent molecules with large binding energies (e.g., \ce{H2O}, SiO, SiS) can be depleted from the gas phase onto the dust \citep{VandeSande2019b}. 
The modelled depletion levels are in line with observations \citep{Decin2010a,Lombaert2013}.
In lower-density outflows, dust-gas interactions do not have a significant influence on the gas phase, as the accretion rate of gas-phase species onto dust is proportional to outflow density.

The grain size distribution (GSD) of AGB dust is also largely unknown.
Observations of SEDs, meteoritic samples, and theoretical studies indicate that AGB dust is large ($a \geq 0.1 \mu$m) but not single sized \cite[e.g.,][]{Groenewegen1997,Gauger1999,Hoppe2000,DellAgli2017,Nanni2018}.
The depletion level of gas-phase species depends on the assumed GSD.
The canonical Mathis–Rumpl–Nordsieck (MRN; \citealt{Mathis1977}) distribution is commonly assumed when modelling CSEs.
\citet{VandeSande2020} allow for an MRN-like distribution, characterised by a minimum and maximum grain size and the slope of the dust grain number density distribution.
Because of degeneracies within the prescription of the GSD, only the average dust grain cross-section can be retrieved from a specific depletion level. 
From observed depletion levels \citep{Bujarrabal1989,Keady1993,Schoier2004,Schoier2006b,Fonfria2008,Decin2010a,Agundez2012,Lombaert2013}, we found cross sections that are larger than that of the canonical MRN distribution, suggesting that the production of large grains appears to be accompanied by more small grains than expected by MRN.
The sizes of these small grains correspond with the size of ultrasmall silicates or PAHs \citep{Li2001b,Tielens2005}.
Note however that only a few depletion levels have been retrieved so far, since retrieving an abundance profile requires observations of a range of high- and low-energy transitions probing the entire outflow and computationally intensive radiative transfer modelling.

Dust-gas chemical models can also hint towards the surface composition of the dust as it enters the ISM.
Interstellar dust is thought to have a layered structure of a core surrounded by inner refractory organic mantle, formed by the photoprocessing of complex ices in the diffuse ISM, topped off by an outer ice mantle \cite[e.g.,][]{Greenberg1999,Jones2013}.
In high-density AGB outflows, gas-phase species are accreted onto the dust and form a physisorbed ice mantle; the chemical evolution of interstellar dust might hence already take place in the AGB phase.
\citet{VandeSande2021} included the photoprocessing of complex volatile ices (containing three or more carbon atoms) into inert refractory organic material and found that the dust in high-density C-rich outflows can have a surface coverage of up to $\sim20\%$.
The refractory organic coverage can be even larger, up to a few monolayers, when including a stellar companion in a clumpy outflow \citep{VandeSande2023}.
The surface coating of AGB dust hence depends on the structure of the outflow from which it originates and whether this structure is caused by a stellar companion.
Such a refractory coating could impact the dust's survival rate and GSD as it enters the ISM \cite[e.g.,][]{Maercker2022}.

\section{Observed chemical complexities}			\label{sect:compl}

The gas-phase chemistry within AGB outflows has been studied observationally for several decades and chemical models are generally successful in reproducing the abundances and abundance profiles retrieved from observations.
Nonetheless, several disagreements between observations and chemical models remain, of which I will discuss a few here.
These complexities help reveal missing chemistry and explain dynamical features.

\subsection{Inner wind}

Thermodynamical equilibrium is a reasonable assumption close to the stellar surface because  of the high densities and temperatures in this region \citep{Tsuji1973,Agundez2020}.
Under these conditions, the strong CO molecular bond would lock up the oxygen in C-rich outflows, and carbon in O-rich outflows.
Nevertheless, \ce{H2O} has been detected close to C-rich stars \citep{Decin2010b,Neufeld2011} and C-bearing molecules such as HCN, CS, and CN have been detected close to O-rich stars \cite[e.g.][]{Omont1993,Bujarrabal1994,Schoier2013}. 
The inner wind abundance of \ce{NH3} is an issue for both C-rich and O-rich stars \cite[e.g.][]{Keady1993,Menten2010,Schoier2011,Wong2018}.
Shocks caused by stellar pulsations take the chemistry in the inner region out of equilibrium.
By including the effects of shocks in a semi-analytical approximation, inner wind chemical models are able to explain observed abundances of most of these ``unexpected'' species, but still have difficulties in explaining the abundances of, e.g., \ce{NH3} and \ce{SO2} \citep{Cherchneff2006,Gobrecht2016}.

These inner wind models assume that the inner region is fully shielded against interstellar UV radiation and hence do not include photochemistry.
In a spherically asymmetric outflow, however, density inhomogeneities could lead to interstellar UV photons reaching the inner wind, liberating C, N, and O by photodissociating parent species.
By allowing a set fraction of interstellar UV photons through, \citet{Agundez2010} were able to abundantly produce unexpected species in C-rich and O-rich inner winds.
In the porosity formalism, interstellar photons still experience extinction, which depends on the specific clumpiness of the outflow.
Only \ce{NH3} shows a significant increase using this approach \citep{VandeSande2018Err}.

In the inner wind of IRC+10216, NaCN, \ce{CH3CN}, \ce{C4H2} are abundantely present \citep{Agundez2015,Quintana-Lacaci2017,Fonfria2018}. 
\citet{Siebert2022} recently added \ce{HC3N} to this list using archival ALMA observations. 
While the presence of these molecules in the inner wind is not unexpected from a thermodynamical equilibrium point of view, their retrieved abundances are too large to be explained by both inner wind models and classical chemical kinetics models of the outer wind.
The companion UV photon model \citep{VandeSande2022} was able to enhance the \ce{HC3N} abundance in the inner wind by including a solar-like companion and adjusting the porosity of the outflow to fit IRC+10216, showing that chemistry can be used to reveal the presence of a stellar companion.

Chemistry can also be used to constrain the orbital parameters of the binary system.
The S-type star W Aql is a known long-period binary. 
\citet{Danilovich2024} found anisotropic emission of SiN and SiC in its outflow, a relic of the solar-like companion's most recent periastron passage.
Combining the companion UV photon chemical model and hydrodynamical modelling, they developed a new astrochemistry-based method to reveal the orbital parameters of a highly eccentric binary orbit.

\subsection{Outer wind}

A long-standing puzzle concerns the distribution of cyanopolyynes (HC$_n$N, with $n = 3,5,7,..$) and hydrocarbon radicals (C$_n$N,  with $n = 2,4,6,...$) in the outer regions of IRC+10216's outflow.
This well-studied outflow contains multiple broken shell-like structures spaced at regular intervals that are thought to be caused by episodes of high mass loss \cite[e.g.,][]{Mauron2000,DeBeck2012,Cernicharo2015}.
The cyanopolyynes and hydrocarbon radicals have their observed emission maxima in one of these shells.
They are formed via UV-driven polymerisation, where photodissociation of the parents \ce{C2H2} and HCN is followed by reactions that build up the carbon chains \cite[e.g.,][]{Agundez2017}.
The observed emission maxima of the cyanopolyynes show a radial sequence, as expected from their formation pathways.
The emission maxima of the hydrocarbon radicals, however, are observed to be cospatial \citep{Guelin1999,Agundez2017,Keller2017}.
Chemical kinetics models have been developed to explain this discrepancy with observations.
\citet{Brown2003} and \citet{Cordiner2009} developed a gas-phase chemical kinetics model that took the effect of the density-enhanced dust shells within the outflow into account. 
These density enhancements resulted in cospatial peaks for both cyanopolyynes and hydrocarbon radicals.
\citet{Agundez2017} assumed a smooth outflow, but lowered the extinction throughout the outflow by a factor of 1.5 to approximate the influence of the clumpy substructures in the dusty shells around IRC+10216. 
This model predicts a radial sequence for both families of molecules.
Therefore, no single model is able to explain the different behaviour of the cyanopolyynes and hydrocarbon radicals, despite IRC+10216 being the most-studied AGB star.
Cyanopolyynes appear to be cospatial around other C-rich stars, though high-resolution observations are needed to better resolve the molecular shells \citep{Unnikrishnan2023}. 

The abundances of S-bearing molecules around O-rich stars depend on mass-loss rate.
\ce{H2S}, CS, and SiS have larger initial abundances as mass-loss rate increases \citep{Danilovich2017,Danilovich2018,Danilovich2019,Massalkhi2019}.
The abundance profile of SO and \ce{SO2} changes with mass-loss rate: lower mass-loss rate outflows show the expected Gaussian decline, which changes to a lower initial abundance followed by a bump in abundance before photodissociation as mass-loss rate increases \citep{Danilovich2019,Danilovich2020,Wallstrom2024}.
In low mass-loss rate outflows, all S appears to be locked up in SO and \ce{SO2}, leading to smaller abundances of the other S-bearing species.
The difference in abundance profile for SO and \ce{SO2} with mass-loss rate cannot be explained using the classical, smooth outflow chemical model.
Including the effects of a clumpy outflow and a companion star can both reproduce the bump in abundance for higher mass-loss rate outflows \citep{Danilovich2020}.
Observations of a combination of molecules are hence necessary to unravel the underlying physics.

\section{Discussion}			\label{sect:discussion}

To help solve the issues of dust formation and observed gas-phase chemical complexities, various needs have to be addressed.
Sects \ref{subsect:disc:obs} and \ref{subsect:disc:model} give a non-comprehensive list of the observational and modelling needs, respectively.

\subsection{Observational needs}		\label{subsect:disc:obs}

To test the validity of a chemical model, it is essential to retrieve not just abundances from molecular lines, but also the abundance profile throughout the outflow.
This is a time-consuming process in several ways: observations of multiple molecular lines covering a range in energy levels are needed to probe the entire outflow, which then need to be modelled using radiative transfer.
Nevertheless, it is time well-spent, as the shape of a retrieved abundance profile can reveal which specific physical and chemical processes are at play.
For example, retrieved depletion levels can help constrain the dust grain-size distribution \citep{VandeSande2021}, while the shape of the abundance profiles of certain species can indicate the presence of a stellar companion \cite[e.g.,][]{Siebert2022,VandeSande2022}.

In order to retrieve abundance profiles via radiative transfer, collisional rates specific to the observed molecules are essential.
\citet{Danilovich2017} highlight the importance of this when retrieving the \ce{H2S} abundance in a sample of outflows.
Two sets of collisional rates exist for \ce{H2S}, but both are scaled \ce{H2O} collisional rates \citep{Dubernet2006,Faure2007,Dubernet2009,Daniel2010,Daniel2011}, neglecting differences in dipole moments and molecular cross sections.
Additionally, the sets of rates are appropriate for different temperature ranges, which are all present within an outflow \citep{vanderTak2011}. 
Using the different sets in an AGB outflow radiative transfer model yields different abundances and sizes of the \ce{H2S} envelope \citep{Danilovich2017}.

When performing observations, it is important to not just target the ``classical'' molecules that are used to retrieve the dynamics of the outflow, such as CO and SiO.
Observations of a well-chosen set of molecules can help determine the stellar or substellar nature of a companion \citep{VandeSande2022,Siebert2022,Danilovich2024} and help constrain the seeds of dust formation \cite[e.g.][]{Decin2017,Danilovich2021}.
Additionally, comparing the molecular composition of different outflows can help disentangle physics and chemistry \citep{Unnikrishnan2023,Wallstrom2024}. 

The \emph{James Webb Space Telescope} offers a new window through which to study the dust.
Tracing the dust's growth and its icy and refractory organic surface composition throughout the outflow will shed a different light on our understanding of its formation and processing.
Combining these data with observations of planetary nebulae and the ISM, projects which are underway, allows for the study of the evolution of the dust as the star evolves off the AGB and the dust enters the ISM.

\subsection{Modelling needs}			\label{subsect:disc:model}

Hydrodynamical models of binary interactions shaping the outflow offer promising results, reproducing some of the intricate observed structures  \cite[e.g.,][]{Kim2012,ElMellah2020,Malfait2021,Maes2021}.
To compare the density distribution computed by these hydrodynamical models to observed molecular emission, chemical models are needed to calculate the molecular abundances, followed by 3D radiative transfer modelling.
It is possible to approximate this for CO, as its molecular extent is mostly determined by photodissociation \cite[e.g.,][]{Mamon1988,Saberi2019,Danilovich2024}. 
For other molecules however, adding an extended, or even complete, chemical network is necessary. 

A 3D snapshot chemical model can be generated by post-processing hydrodynamical model output, as is done for protoplanetary disks \cite[e.g.,][]{Walsh2014}.
To develop a 3D time-dependent chemical model, it is essential to include the chemistry in a hydrodynamical model.
However, calculating the chemical kinetics at each step in a hydrodynamical model would add a prohibitively large amount of computation time.
This can be circumvented by using a reduced reaction network \cite[e.g.,][]{Boulangier2019a}, but this drastically limits the versatility of the chemical model.
An alternative approach to speeding chemistry is the use of machine learning techniques to emulate the chemistry, rather than solving the network. 
These chemical emulators are so far available for interstellar chemistry only and again use a reduced reaction network \citep[e.g.,][]{Grassi2011,deMijolla2019,Holdship2021}.
A chemical emulator geared towards AGB outflows that includes a complete reaction network is currently being developed.
The emulator takes the density, temperature, and radiation output of the hydrodynamical model and abundances at a specific time step and uses this to predict the abundances for the next time step, reducing the computation time by several orders of magnitude (Maes et al., in prep.). 
Such an emulator would allow for the development of a computationally feasible 3D time-dependent hydrochemical model.

\section*{Acknowledgements}			
\noindent
I would like to thank the organisers for inviting me to participate in this conference.
At the time, I was supported by the European Union's Horizon 2020 research and innovation programme under the Marie Skłodowska-Curie grant agreement No 882991. I am now grateful for the Oort Fellowship at Leiden Observatory.

\bibliographystyle{iaulike}
\bibliography{chemistry}

\end{document}